\documentclass[10pt]{iopart}
\usepackage{graphicx}
\usepackage{color}

\begin{document}
\title[A modular, extendible and field-tolerant multichannel vector magnetometer based on current sensor SQUIDs]{A modular, extendible and field-tolerant multichannel vector magnetometer based on current sensor SQUIDs}

\author{J-H Storm$^1$, D. Drung$^1$, 
M. Burghoff$^1$ and R. K{\"o}rber$^1$}

\address{$^1$ Physikalisch-Technische Bundesanstalt (PTB), Abbestrasse 2-12, D-10587 Berlin}
\ead{Jan-Hendrik.Storm@ptb.de}

\vspace{10pt}
\begin{indented}
\item[\today]
\end{indented}

\begin{abstract} 
We present the prototype module of our extendible and robust multichannel SQUID magnetometer system. A large multi-module arrangement can be implemented by using up to 7 modules. The system is intended for high-precision measurements of biomagnetism and spin precession. Further demanding applications are magnetorelaxometry and ultra-low-field nuclear magnetic resonance (ULF NMR), where pulsed magnetic fields of up to 100\,mT are typically applied. The system is operated inside the Berlin Magnetically Shielded Room (BMSR-2) and equipped with 18 magnetometers consisting of niobium (Nb) wire-wound pick-up coils. A total of 16 small pick-up coils with 17.1\,mm diameter form a regular grid with individual channels arranged to ensure system sensitivity covers all three orthogonal spatial directions. Two large hexagonal pick-up coils with an equivalent diameter of 74.5\,mm sensitive in z-direction surround the grid at two different heights and are suitable for the detection of deep sources. Each pick-up coil is connected to the input of a thin-film Nb SQUID current sensor via a detachable superconducting contact. The SQUIDs are equipped with integrated input current limiters. Feedback into the pick-up coils is employed to minimize crosstalk between channels. The current sensor chip package includes a superconducting shield of Nb. The field distortion of the prototype and a multi-module arrangement was analysed by numerical simulation. The measured noise of the small magnetometers was between 0.6 and 1.5\,fT/$\textrm{Hz}^{1/2}$, and well below 1\,fT/$\textrm{Hz}^{1/2}$ for the large ones. Using a software gradiometer, we achieved a minimum noise level of 0.54\,fT/$\textrm{Hz}^{1/2}$. We performed ULF NMR experiments, verifying the system's robustness against pulsed fields, and magnetoencephalographgy (MEG) on somatosensory evoked neuronal activity. The low noise performance of our 18-channel prototype enabled the detection of high-frequency components at around 1\,kHz by MEG.
\end{abstract}

\vspace{2pc}
\noindent{\it Keywords: SQUID, multichannel SQUID system, Ultra-low-field NMR, MEG}\\
\submitto{\SUST}
\maketitle
\ioptwocol	

\section{\label{sec:introduction}INTRODUCTION}
Large multichannel SQUID systems have been built from the late 1980's for magetoencephalography (MEG) and magnetocardiography (MCG) measurements in the field of biomagnetism~\cite{Foglietti1992}, allowing the estimation of the location of ionic currents from their magnetic field distribution. Further development has led to the commercialisation of SQUID systems for MEG and MCG~\cite{Vrba2006}. 
\par A more recent application of SQUIDs is ultra-low-field magnetic resonance (ULF MR). Here, the volume of interest is first prepolarised by a field of up to 100 mT and then the SQUID sensor detects the MR signal at a much lower field (typically tens to hundreds of $\mu$T). Usually, such systems are based on coupling a superconducting pick-up coil inductively to a SQUID and can be used for both MEG and ULF MR~\cite{Clarke2007,ULFNMR2014}. In the field of biomagnetism, potential applications include direct neuronal current imaging (NCI) and the combination of ULF MRI and MEG in one instrument. The advantages of combining those two modalities lies in the improvement of source localization accuracy by restricting the available source space by anatomical knowledge. In addition, co-registration errors can be minimized thanks to the common sensor array. Several field-tolerant multichannel systems were built utilizing SQUID current sensors coupled to a pick-up coil~\cite{Zotev2007,Vesanen2012}. Those systems are usually operated in moderately shielded rooms with two layers of mu-metal necessitating the use of axial or planar gradiometric pick-up coil designs to sufficiently suppress ambient low-frequency noise.
\par In this paper, we describe the design, system parameters and demonstrator experiments of an 18-channel prototype of our new modular and field-tolerant multichannel SQUID system. The concept is based on our single-channel system which was used both for ultra-sensitive MEG~\cite{Fedele2015} and ULF-MR measurements~\cite{Koerber2013}. The new multichannel system is to be used as a multipurpose device for MEG, ULF MRI and magnetorelaxometry (MRX) and will be operated in the heavily shielded room BMSR-2 of PTB~\cite{Bork2000} with seven layers of mu-metal and one additional aluminium eddy-current shield. The superior low-frequency shielding performance of the BMSR-2 allows the use of wire-wound magnetometer pick-up coils. We introduce overlapping pick-up coils of different sizes that improve the signal-to-noise ratio (SNR) depending on the depth of the sources with respect to the sensor coil.

\section{\label{sec:module and system design}MODULE AND SYSTEM DESIGN}
In this section, we describe in detail the design of the prototype module. We chose a hybrid concept, in which an array of closely spaced small coils is surrounded by a single large coil. A small grid spacing is necessary in order to scan the field profile with sufficient spatial frequencies to localise cortical sources in MEG~\cite{Ahonen1993a}. In addition, as will be shown below, a large coil is more sensitive to deep sources and can possibly perform better in terms of SNR, making it suitable for the investigation of high-frequency oscillations as observed in MEG, for instance.

\subsection{\label{subsec:hardware}Hardware}
\begin{figure}[t]
\centering
\includegraphics[width=0.5\textwidth]{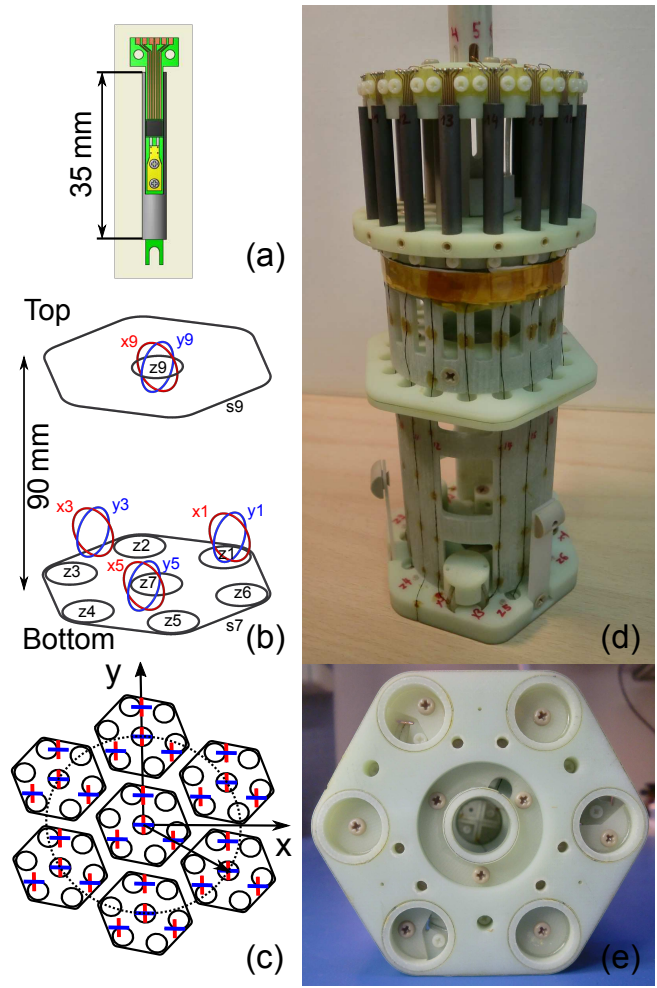}
\caption{\label{Fig:system design}Schematic of the SQUID packaging (a). Sensor layout of the prototype module (b). Top view of 7 module arrangement showing the regular sensor arrays in x, y, and z-direction (c). Picture of prototype module (d). Bottom view depicting the seven small $z$-loops surrounded by the large hexagonal-shaped $z$-loop (e).}
\end{figure}

The SQUID mounting is shown in \fref{Fig:system design}(a). It is glued onto a printed-circuit-board (PCB) carrier with the SQUID positioned centrally inside a 5\,mm diameter, open-ended Nb tube to shield it from external fields. The pick-up coils are connected to the SQUID input coil via a detachable contact, which is fixed via two plastic screws. The SQUID sensors are arranged in a circular fashion at a SQUID plate which can be used as a basic module for any pick-up coil geometry.
\par The multichannel system contains 18 magnetometers as shown in \fref{Fig:system design}(d). The magnetometers are designed with Niobium wire-wound flux antennas and SQUID current sensors from the sensor family PTB-C7, which are based on the double-transformer scheme presented in~\cite{Drung2007}. For best possible impedance matching between SQUID and the pick-up coils, we use three types of transformers with input inductances of 65\,nH, 150\,nH and 400\,nH respectively. Each of the $3.3\times3.3\,\textrm{mm}^{2}$ current sensor chips is equipped with a feedback transformer inductively coupled to the input coil~\cite{terBrake1986}. This allows for cancellation of the signal current in the pick-up coils to minimise crosstalk between the neighboring pick-up loops. To avoid large currents in the input circuit, an on-chip current limiter in series with the input coil is realized. This consists of a series array of 16 unshunted 20\,pH SQUIDs and allows for control of the input current between $\approx 2\,\mu \textrm{A} - 20\,\mu \textrm{A}$ via a separate flux control line. 
\par Pick-up coils with two different sizes, located at two levels separated by a baseline of 90\,mm are used. The nomenclature for the 18 channels is given in \fref{Fig:system design}(b). In the bottom level, for the z-component of the magnetic flux density vector $\bi{B}$, a sensor array is formed by 7 circular coils with 17.1\,mm in diameter arranged in a hexagonal grid with a centre-to-centre distance of 30\,mm. In addition, a large hexagonal pick-up coil with rounded corners surrounds the 7 small coils. The distance between the parallel sides is 71.1\,mm and the outside diameter is 79.1\,mm. It has an equivalent diameter of 74.5\,mm. Reference magnetometers for $B_z$ are mounted in the upper level consisting of one pick-up coil with 17.1\,mm diameter and one with 74.5\,mm equivalent diameter, respectively. The transverse components $B_x$ and $B_y$ of $\bi{B}$ are measured with three 17.1\,mm coil pairs in the lower and one for reference in the upper level of the module. In the lower level, the distance of the  centres of the xy-coils to the bottom level amounts to 11.5\,mm. The coordinate system of the xy-coil pairs is rotated with respect to the z-coil system by $10.89^°$ degree. This facilitates a regular hexagonal grid also for the xy-coil pairs in an extended multiple modules arrangement with a centre-to-centre distance of 45.8\,mm in the bottom level as can be seen in \fref{Fig:system design}(c). The centre of the Nb shield is located 62\,mm above the top level and 152 mm above the bottom level.

\subsection{\label{subsec:Numerical field simulations}Numerical field simulations}

For high-precision magnetic measurements, a detailed knowledge and control of the environmental magnetic field is mandatory. The BMSR-2 offers an excellent platform for such measurements thanks to its superior shielding performance~\cite{Bork2000}. In the central volume of $0.5\times0.5\times0.5\,\textrm{m}^3$ the residual field is about 500 pT with a gradient of 500\,pT/m. This in turn entails high demands on the experimental setup to be used in such an environment.
\par
\begin{figure}
\centering
\includegraphics[width=0.5\textwidth]{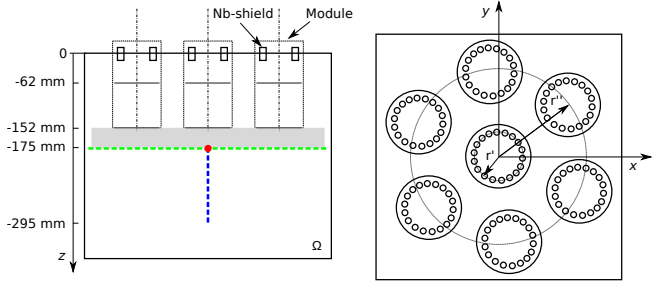}
\caption{\label{Fig:FEM_schema}Left: Front view of setup used for FEM-calculation. dark gray lines indicate pick-up loop levels. The red dot and the dashed green lines indicate the position of the line plots in \fref{Fig:field_dist_xy}, the blue one that of \fref{Fig:field_dist_z}. Right: Top view of the Nb-shield arrangement; $r'=29.5\,\textrm{mm}$, $r''=79.5\,\textrm{mm}$.}
\end{figure}

 The dominant influence of the multichannel system to the magnetic environment results from field distortions caused by the superconducting shields in the presence of static magnetic fields. In order to evaluate this as well as the shielding factor inside the Nb tubes, we solved the magnetostatic problem by making use of the finite element method (FEM). The geometry of the model is depicted in \fref{Fig:FEM_schema}; therein the exterior space is described by stationary Maxwell equations
\begin{eqnarray}
  \nabla \times \bi{H}=\bi{j} \label{eq.1} \\
  \nabla \cdot \bi{B}=0 \label{eq.2}
\end{eqnarray}
where $\bi{H}=\bi{B}/\mu$ is the magnetic field, $\mu$ the magnetic permeability and $\bi{j}$ the current density. For simplification of the model, the superconducting tubes were excluded and replaced by the boundary condition $\bi{n} \cdot \bi{B}=0$, where $\bi{n}$ is the normal vector on the tube surface. By making use of the model symmetry at $z=0$, the current density vanishes in the half space $\Omega$. Therefore the magnetic field vector can be obtained from a scalar potential:
\begin{eqnarray}
	\bi{H}=-\nabla \Psi . \label{eq.3}	
\end{eqnarray}
Inserting \eref{eq.3} into \eref{eq.2} gives the system descriptive equation:
\begin{eqnarray}
	\nabla \cdot \left( \mu \nabla \Psi \right)=0 \label{eq.4}
\end{eqnarray} 
If we assume a constant $\mu$ in $\Omega$, equation~\eref{eq.4} reduces to the Laplace equation, which can be solved very efficiently by FEM even for high spatial resolution in 3D-space. For the field calculations we use the commercial software package COMSOL Multiphysics.

To verify our numerical model, we simulated a single tube and compared the results with theoretical calculations for the volume inside the tube given in \cite{Claycomb1999}. Furthermore, from these results we calculated the minimum shielding factor $S=B/B_0$ in the area were the SQUID-chip is located. For an external field $B_0$ along the tube axis, we found $S=5\times10^{-12}$ and for a transverse field $B_0$ perpendicular to the SQUID plane $S=3\times 10^{-6}$. Together with the gradiometric design of the SQUID sensors this is sufficient for almost all applications of the multichannel system.

\par
\begin{figure}[hbtp]
\centering
\includegraphics[width=0.5\textwidth]{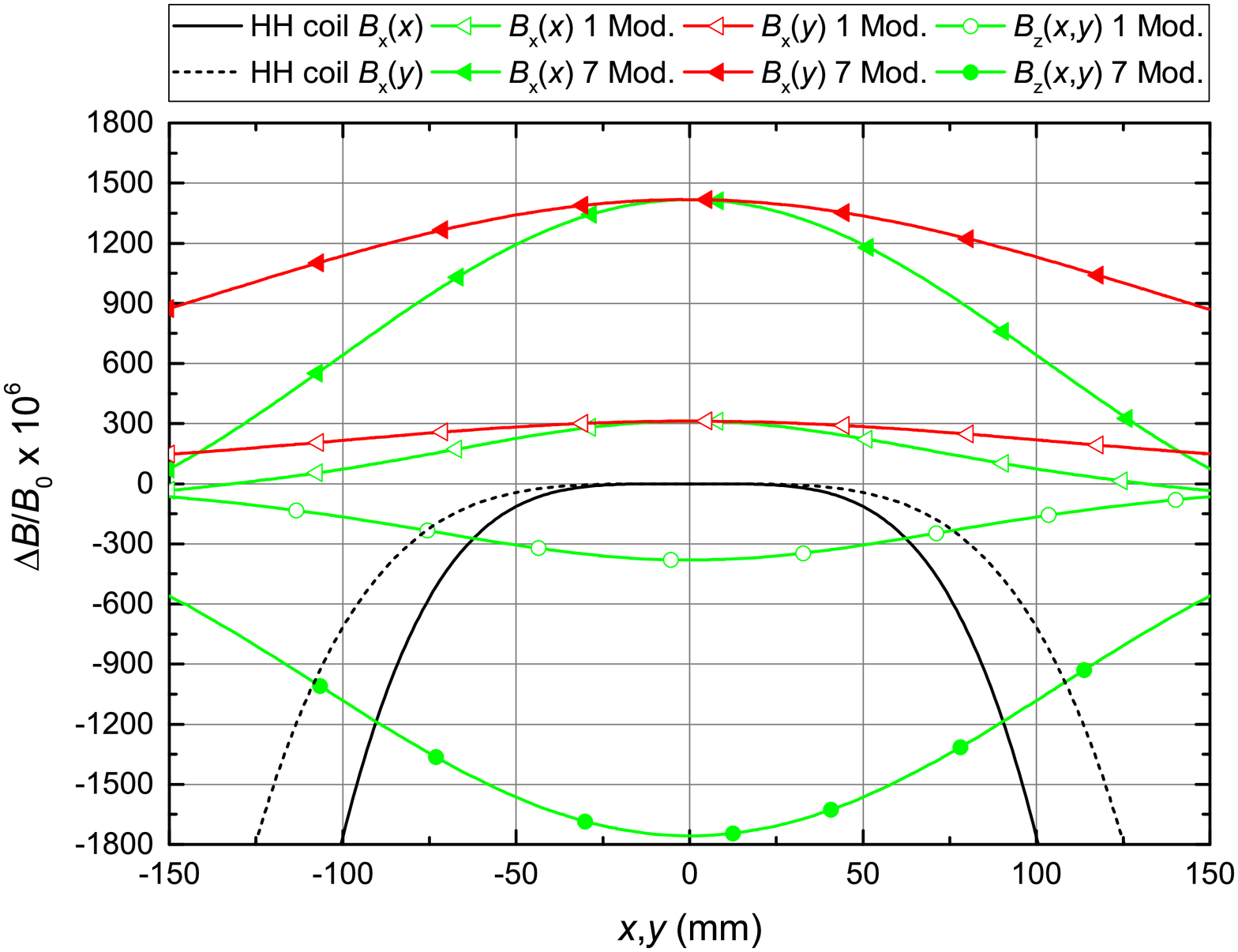}
\caption{\label{Fig:field_dist_xy}Horizontal distortion of a homogeneous background magnetic field compared with the field profile of an ideal Helmholtz coil system with 1\,m in diameter.}
\end{figure}

\begin{figure}[hbtp]
\centering
\includegraphics[width=0.5\textwidth]{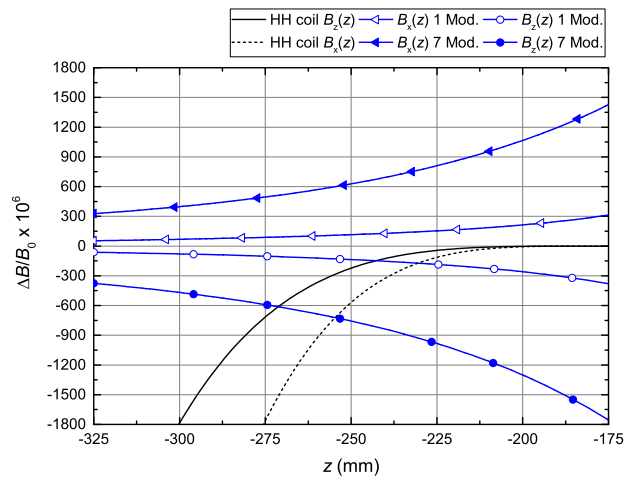}
\caption{\label{Fig:field_dist_z}Vertical distortion of a homogeneous background magnetic field compared with the field profile of an ideal Helmholtz coil system with 1\,m in diameter.}
\end{figure}

 In the next step, we simulated the 18 shield tubes of the module as well as the arrangement of seven modules. The results are depicted in \fref{Fig:field_dist_xy} and \ref{Fig:field_dist_z} for two experimental situations in each case. In the first case, the external field $B_{0}$ is parallel to the x-axis; in the second case, the field is aligned along the z-axis. For the illustration, the normalized deviation $(B-B_0)/B_0$ of the vector field component which is pointing in the direction of the applied field is shown. The three colours indicating cut lines of the vector field lying on the x-, y- and z-axis with their origin $175\,\textrm{mm}$ below the centre of the shield tubes. This corresponds to the outside surface of the dewar bottom as depicted in \fref{Fig:FEM_schema}. For comparison the deviation of an ideal Helmholtz coil with $1\,\textrm{m}$ diameter is shown as well. Because of the rotational symmetry of the Helmholtz coil, in \fref{Fig:field_dist_xy} only the case for a $B_{0}$ along the x-direction is given.

\par 
When we consider a small sample volume of $10\times10\times10\,\textrm{mm}^3$ central below the module and an external field in x-direction, e.g., for NMR spectroscopy, the field distortions from the shields over the sample size are $1.5\times10^{-6}$, $0.5\times10^{-6}$ and $45\times10^{-6}$ for the x-, y- and z-direction respectively. Compared with the ideal Helmholtz coil, these are two orders of magnitude worse for the xy-direction and three orders of magnitude for the z-direction. However, for a realistic estimation, imperfections in the geometry of the Helmholtz coil should be taken into account. For example, a displacement of $2\,\textrm{mm}$ of one coil arm in the axial direction causes the inhomogeneity to increase by two orders of magnitude and thereby to become comparable to the shield distortions for transverse fields components. Considering the z-direction, the effect of the shields is still a factor of 30 higher. On the other hand, as the sample size increases the Helmholtz coil inhomogeneity worsens more quickly in comparison to the shield distortions. 

\par We have also calculated the imbalance of the magnetometers in the upper and lower planes caused by the shields. Results for the centrally arranged small and the large magnetometers in the center module as well as for the rim modules are given in table~\ref{Tab:imbalance}. The calculated values are of the same order of magnitude as the geometrical imbalance of a wire wound gradiometer. Therefore this influence of the shield tubes is acceptable.
\begin{table}
\caption{\label{Tab:imbalance}Imbalance between the upper and lower pick-up coils  }
\begin{indented}
\item[] \begin{tabular}{@{}lllll}
\br
Field &centre module & & rim module\\
 &z7/z9 &s7/s9 &z7/z9 &s7/s9 \\
\mr
$B_z$&$4.75\rme^{-3}$&$5.27\rme^{-3}$&$3.99\rme^{-3}$&$3.9\rme^{-3}$\\
\br
\end{tabular}
\end{indented}
\end{table}

\subsection{\label{subsec:noise after pulsing}Excess low-frequency noise after pulsing}

The system is to be used as a multi-purpose device and, amongst other, for applications where magnetic field pulses in the range of up to 100\,mT are required. We chose Nb for the pick-up coil as it has a high critical temperature, is widely available as insulated wires and is easy to handle during coil fabrication. More importantly, high-purity Nb has one of the highest lower critical fields $H_{\textrm{c}1}$ among the known type-II superconductors. The exposure of the pick-up coil wire to pulsed fields leads to trapped flux if $H_{\textrm{c}1}$ is exceeded. In real samples this occurs already at a lower field as the transition from the Meissner state to the mixed state is usually broadened. We denote the field at which flux penetration starts as $H_{\textrm{c}1^\prime}$. After field removal, the flux will rearrange and thereby introduce a random telegraph signal manifesting itself as excess low-frequency noise~\cite{Hilschenz2013,Matlashov2015,Luomahaara2011}. Apart from this additional noise, field distortions due to the trapped flux are also observed ~\cite{Hwang2014}.
 
\par
\begin{figure}
\centering
\includegraphics[width=0.475\textwidth]{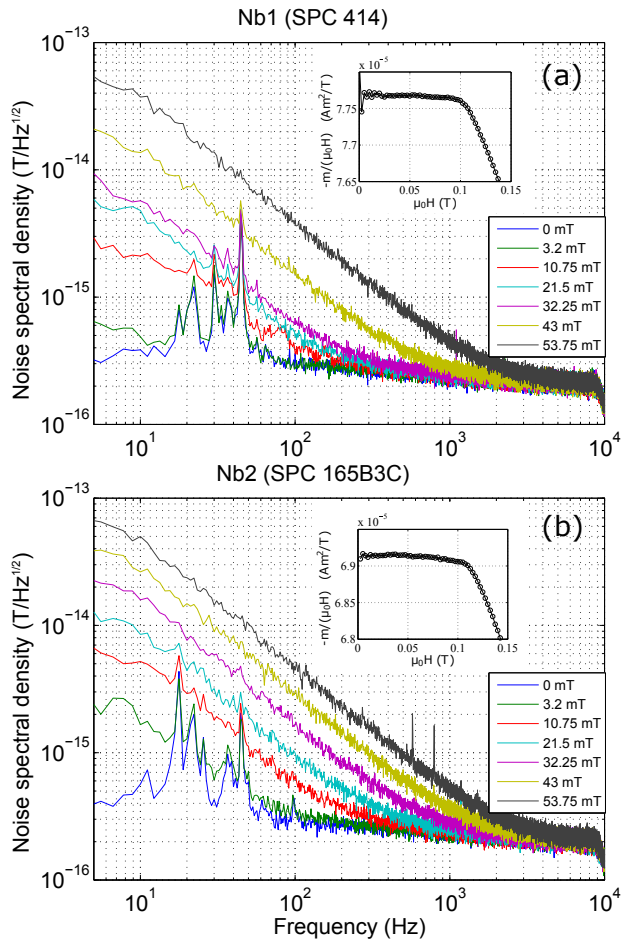}
\caption{\label{Fig:pulsing}Noise after pulsing of test gradiometers made from Nb1 (a) and  Nb2 (b) of niobium wire. The insets show the determination of $H_{\textrm{c}1^\prime}$ at which flux starts to penetrate the wire as determined with an MPMS system.}
\end{figure}

 In order to assess the suitability of various Nb batches of different origin, we determined the critical field $H_{\textrm{c}1^\prime}$ of four different wire samples with diameters ranging from 50.8~$\mu$m to 125~$\mu$m. The field dependence of the static magnetic moment $m$ of approximately 10\,mm-long wire samples at 4.2\,K was measured with a commercial magnetic properties measurement system (MPMS, Quantum Design, San Diego, CA). As shown in the insets of \fref{Fig:pulsing}, we plot $m$ divided by the applied field $\mu_{0}H$ in order to extract $H_{\textrm{c}1^\prime}$, the field at which $m/(\mu_{0}H)$ starts to deviate from a constant value. As the field was applied parallel to the wire axes, demagnetising effects can be neglected. We obtained a wide span for $\mu_{0}H_{\textrm{c}1^\prime}$ ranging from 67~mT to about 100~mT, see table I in reference~\cite{Hwang2014}. A possible explanation for this marked difference in $H_{\textrm{c}1^\prime}$ could be varying amounts of impurities within the separate samples. The batches Nb1 (SPC 414) and Nb2 (SPC 165B3C), both of nominal diameter 101.6 $\mu$m, showed the maximum value with $\mu_{0}H_{\textrm{c}1^\prime}\approx100$\,mT.

\par In order to choose the best possible wire from the available batches, we built small test gradiometers from batches Nb1 and Nb2 and exposed them to pulsed fields. We limited this investigation to Nb1 and Nb2 as it has been shown that pick-up coils made from niobium-titanium with a much lower $\mu_{0}H_{\textrm{c}1^\prime}$ of about 28~mT show significantly more excess low-frequency noise~\cite{Hilschenz2013}. Each axial gradiometer with diameter 41.5\,mm and baseline 4\,mm was wound on a glass-fibre reinforced plastic (GRP) former. It was exposed to pulsed fields oriented along the gradiometer axis in ascending order from 0 to 53.75\,mT after zero-field cooling. This ensured that initially the wire did not contain trapped flux. We calculate the magnetic flux density noise ${S}_{B}^{1/2}$ 100 ms after turn off.

\par Nominally identical wires have different behaviour regarding excess low-frequency noise after pulsing as can be seen in \fref{Fig:pulsing}. From the MPMS measurements, we expect flux penetration at a field of about 50\,mT due to demagnetising effects within the wire for a perpendicularly applied field. For Nb2, a steady increase of the excess low-frequency noise with increasing pulsed fields is observed. At 100\,Hz and for 32.25\,mT the noise amounts to 1.7\,fT/$\textrm{Hz}^{1/2}$. For Nb1, the increase is less pronounced for pulsed fields up to 32.25\,mT at which the noise is 0.65\,fT/$\textrm{Hz}^{1/2}$. Then, an abrupt increase in low-frequency excess noise is observed at and beyond 43\,mT, approximately consistent with the expected threshold of $\mu_{0}H_{\textrm{c}1^\prime}$/2. In fact, at 53.75\,mT the excess noise for both wires is almost equal. Nevertheless, excess noise is already seen at lower fields for both lots indicating an unidentified additional source. 

\par The pick-up coils of the prototype module were wound of Nb1 and at present it is not clear why Nb2 shows significantly more excess noise at lower pulsed fields. Further experiments rule out the insulation or the former material as the origin. In addition, the MPMS measurement is not sensitive enough to enable a screening of wires without the need to actually perform pulsed field experiments. A more detailed study, preferably with a larger number of wire batches, is necessary to investigate and determine the source of this excess low frequency noise observed at fields smaller than $H_{\textrm{c}1^\prime}$.

\section{\label{sec:experimental results}EXPERIMENTAL RESULTS AND DISCUSSION}
In this section, the main parameters of the prototype module and the demonstrator experiments are described and discussed. The module was operated in a low-noise liquid helium dewar (Fujihira Co., Ltd., Tsukuba, Japan) with a flat bottom of 250\,mm diameter and a nominal warm-cold distance of 24\,mm. Positioned inside the BMSR-2, the sensors were read out by a XXF-1 SQUID-electronics from Magnicon~\cite{Magnicon}. Further details on the ULF NMR and MEG experiments are given in the respective sections below.

\subsection{\label{subsec:system characteristics}System characteristics}

For the baseline characterisation of the prototype module, the field sensitivity $B_{\Phi}$ and the flux density noise of each sensor were experimentally determined. The field sensitivity is given by $B_{\Phi}=L_{tot}/(M_{in}~A_p)$~\cite{Clarke2004}, where $L_{tot}$ is the total inductance of the input circuit, $M_{in}$ is the mutual inductance between input coil and SQUID and $A_{p}$ is the field sensitive area of the pick-up loop. $L_{tot}$ can be determined from two measurements of the mutual inductance $M_f$ between feedback coil and SQUID. In one case the pick-up coil is connected to the input coil, in the other case a low inductive short instead. $L_{tot}=L_p+L_{str}+L_{in}$ is then calculated from the ratio $M_{f,short}\,/\,M_{f,pickup}=(L_p+L_{str}+L_{in})/(L_{in}+L_{short})$ where $L_p$, $L_{str}$, $L_{in}$ and $L_{short}$ are the inductances of the pick-up loop, the interconnection line, the input coil and the short, respectively. $L_{in}$ was determined independently by a separate measurement. Neglecting the minute contribution from $L_{short}$ we find for the small bottom magnetometers a mean $B_{\Phi}$ of 890\,pT/$\Phi_0$ and 88.3\,pT/$\Phi_0$ for the large one.
Here $\Phi_0$ is the magnetic flux quantum. In the upper plane $L_{str}$ is reduced, which leads to a mean $B_{\Phi}$ of 780 and 86.4\,pT/$\Phi_0$ for the small and the large coil, respectively.

\par
\begin{figure}[t]
\centering
\includegraphics[width=0.5\textwidth]{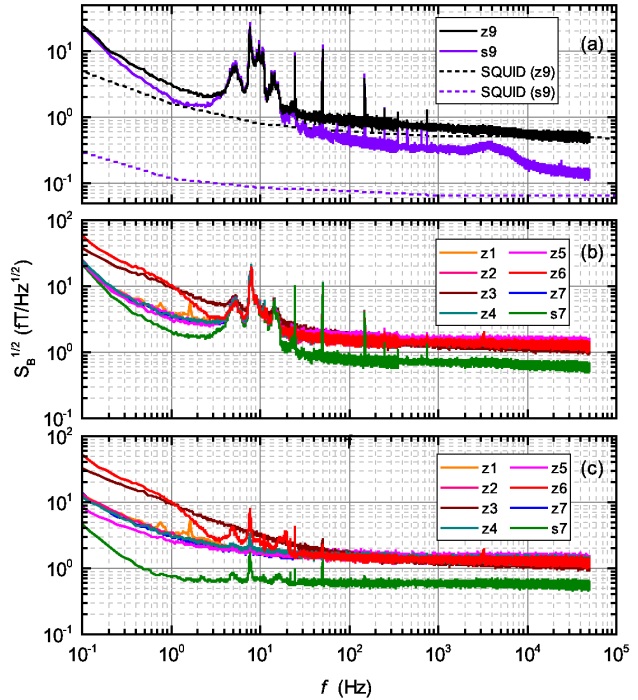}
\caption{\label{Fig:Noise_z_ch}Noise spectral density of the channels $z9$ and $s9$ (a) and of the bottom z-channels $z1$--$z7$ and $s7$ (b). Noise spectral density of the $z$-channels after implementing software gradiometers using $s9$ as the reference magnetometer (c).}
\end{figure}

 Considering the field sensitivities above, together with the intrinsic white flux noise ${S}_{\Phi}^{1/2} \approx 0.7\mu \Phi_0/\textrm{Hz}^{1/2}$ of the SQUID sensors, we obtain a white field noise between 0.54\,-\,0.62\,$\textrm{fT}/\textrm{Hz}^{1/2}$ for the small coils and $\approx 0.061\,\textrm{fT}/\textrm{Hz}^{1/2}$ for the large coils. The intrinsic sensor noise for the upper z-channels is given in~\fref{Fig:Noise_z_ch}~(a) (dotted lines). Comparing this with the measured field noise in the prototype setup, it can be seen that the small magnetometer $z9$ reaches its intrinsic value above 10\,kHz. For the lower frequencies, the measured noise is increased by ambient noise. The large magnetometer $s9$ is dominated by ambient noise over the full measured frequency range.

\par
\begin{figure}[t]
\centering
\includegraphics[width=0.5\textwidth]{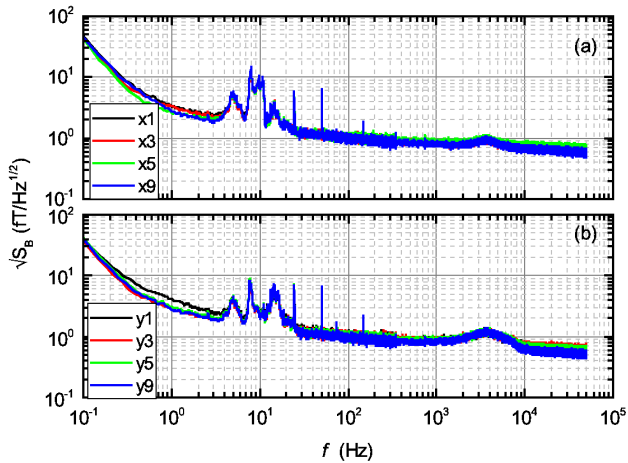}
\caption{\label{Fig:Noise_xy_ch}Noise spectral density for x-channels (a) and y-channels (b).}
\end{figure}

 The field noise of the bottom $z$-channel magnetometer is depicted in \fref{Fig:Noise_z_ch}(b). In order to reduce background field fluctuations, we implemented synthetic gradiometers~\cite{Clarke2004} using $s9$ as the reference (\fref{Fig:Noise_z_ch}(c)). For the small $z$-gradiometers in the bottom plane, we found a white noise level of about 1.28\,$\textrm{fT}/\textrm{Hz}^{1/2}$. For the large hexagonal-shaped coil $s7$ it is 0.61\,$\textrm{fT}/\textrm{Hz}^{1/2}$ for the magnetometer and 0.54\,$\textrm{fT}/\textrm{Hz}^{1/2}$ in the gradiometer case, respectively. The uncorrelated noise of $s9$ is significantly smaller than that of the bottom sensors as we observe a further improvement of the white noise level after gradiometer implentation. The minimum noise level for both, the small and the large bottom channels is limited by thermal noise emitted from the dewar bottom. Correlation of the thermal noise within the pick-up coil results in the measured noise level being dependent on the coil size~\cite{Nenonen1996}. For our geometry, we expect from the calculations in \cite{Nenonen1996} a noise reduction of about a factor of two for the large coil compared to the small ones, which is in reasonable agreement with the measured data. At low frequencies, the SQUID of channel $z$3 has pronounced intrinsic  1/f-noise. The additional disturbance around 10\,Hz in channels $z1$, $z6$ and $s7$ still observed after calculating gradiometric signals arise from a magnetic impurity within the dewar bottom. This was verified by a rotation of the module with respect to the dewar after which other channels showed this additional noise components.
In \fref{Fig:Noise_xy_ch}, we show the field noise for the $x$ and $y$ sensors. Above 100\,Hz, the sensors have a noise below 1\,$\textrm{fT}/\textrm{Hz}^{1/2}$. The pronounced broad peak at around 4\,kHz (also seen in $s9$) is likely caused by interference from the electronics cabling.

\par If SQUID noise is the only noise component, the large coil will always perform better in terms of ${S}_{B}^{1/2}$ compared to a combination of the small ones. For uncorrelated noise, ${S}_{B}^{1/2}$ will be reduced by $\sqrt{7}$. Taking 0.54\,fT/$\textrm{Hz}^{1/2}$ for a single small coil this results in a final ${S}_{B}^{1/2}$ of 0.20\,fT/$\textrm{Hz}^{1/2}$, about a factor of three larger than the intrinsic ${S}_{B}^{1/2}$ of 0.061\,fT/$\textrm{Hz}^{1/2}$ of the large coil.
\par In terms of SNR, the better performance of the large coil is only applicable if the magnetic field of the signal is homogeneous over the pick-up coil array. In biomagnetic or ULF NMR applications, this is, however, rarely the case and in order to optimise the pick-up coil geometry with respect to maximum SNR, one has to perform a detailed analysis for a particular experiment which takes into account the field-profile of the signal and the various noise contributions, see for example~\cite{Matlashov2012}. 
\par A simple example illustrates this point. Assume a magnetic dipole pointing along $z$ and located centrally under a circular pick-up coil at a depth $t$. In this case, the optimum diameter $d$ for detecting $B_z$ is $2\sqrt{2}t$ as the maximum flux threading the pick-up coil is collected. For diameters larger than $d$, $B_z$ changes sign so that even for SQUID-limited noise the SNR decreases. If additional ambient noise is present, $d$ will be somewhat smaller as one is collecting more noise in comparison to signal flux.

\subsection{\label{subsec:ultra-low-field nmr}Ultra-low-field NMR}

In order to verify the robustness against pulsed fields, ULF NMR was carried out on a sample of distilled water.
A solenoid with inner diameter of 28\,mm, height of 35\,mm and a $B/I$ of 5.0 mT/A generated the polarising field in the vertical direction. The sample fitted snugly in the polarising coil and and was located centrally below the module. The transverse detection field was generated by a Helmholtz coil with diameter of 1\,m and a $B/I$ of 5.69\,$\mu$T/A. The polarising current was supplied by 12\,V batteries switched by solid-state relays. With the resistive load of the coils, the current amounted to 7\,A giving a polarizing fields of 35\,mT at the coil centre. Data were taken for detection fields up to 2.56\,$\mu$T in 9 steps.
\par In these experiments, the noise was limited by leakage current noise from the polarising coil. Therefore, the maximum SNR is achieved for the sensor which covers most of the sample, in this case $z7$. The large hexagonal sensor $s7$ performs worse as the field is not constant over the large coil and the effective field is therefore reduced. 
\par
\begin{figure} [t]
\includegraphics[width=0.475\textwidth]{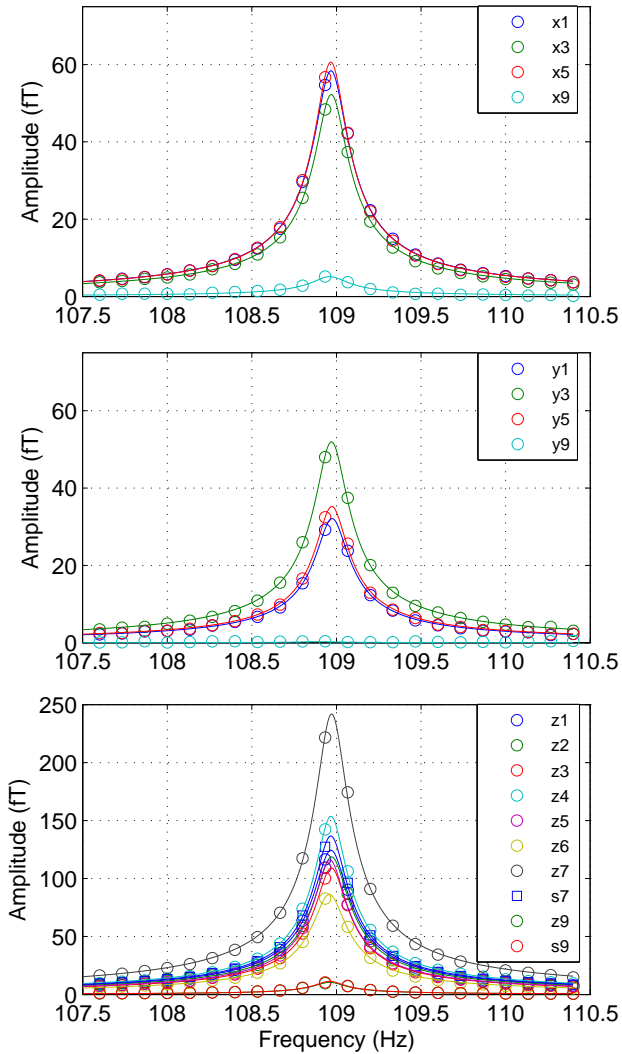}
\caption{\label{Fig:NMR}NMR amplitude spectra for distilled water at 2.56\,$\mu$T.}
\end{figure}

 In \fref{Fig:NMR}, the amplitude spectra are shown for a detection field of 2.56\,$\mu$T. At 109 Hz, the average line-width is about 180\,mHz, corresponding to an effective spin-spin relaxation time $T^{*}_{2}$ of 1.75\,s. The line-width increases with the detection field strength and a linear fit to the results obtained with sensor $z7$ describes the data well. We find an intrinsic line-width of $\beta=(157\pm3)$\,mHz, in agreement with earlier work, but we cannot distinguish between contributions to the line-broadening arising from magnet inhomogeneities, dispersion of $T_{2}$ and influences of the $p$H-value of the sample, see~\cite{Hartwig2011} for details. Assuming inhomogeneous broadening only, we obtain a magnet inhomogeneity of $\alpha=(267\pm37)$\,ppm.

\subsection{\label{subsec:meg}Magnetoencephalography --- MEG}

MEG of somatosensory evoked fields (SEFs) was measured on three healthy volunteers. Electrostimulation of the right median nerve above motor threshold with a square current pulse of 200 $\mu$s duration and a repetition frequency of 9\,Hz was applied to evoke neuronal fields in the contra-lateral somatosensory cortex. The dewar containing the prototype SQUID system was positioned over the left cortical hemisphere with the central sensor (z7) tangentially above the midpoint between C3 and T3 according to the 10-20 convention for EEG electrodes~\cite{Niedermeyer2004}. A session lasted for 30\,min resulting in 16200 stimuli. The inter-stimulus epochs were averaged to reveal the SEFs. 
\par
\begin{figure*}[ht]
\includegraphics[width=0.95\textwidth]{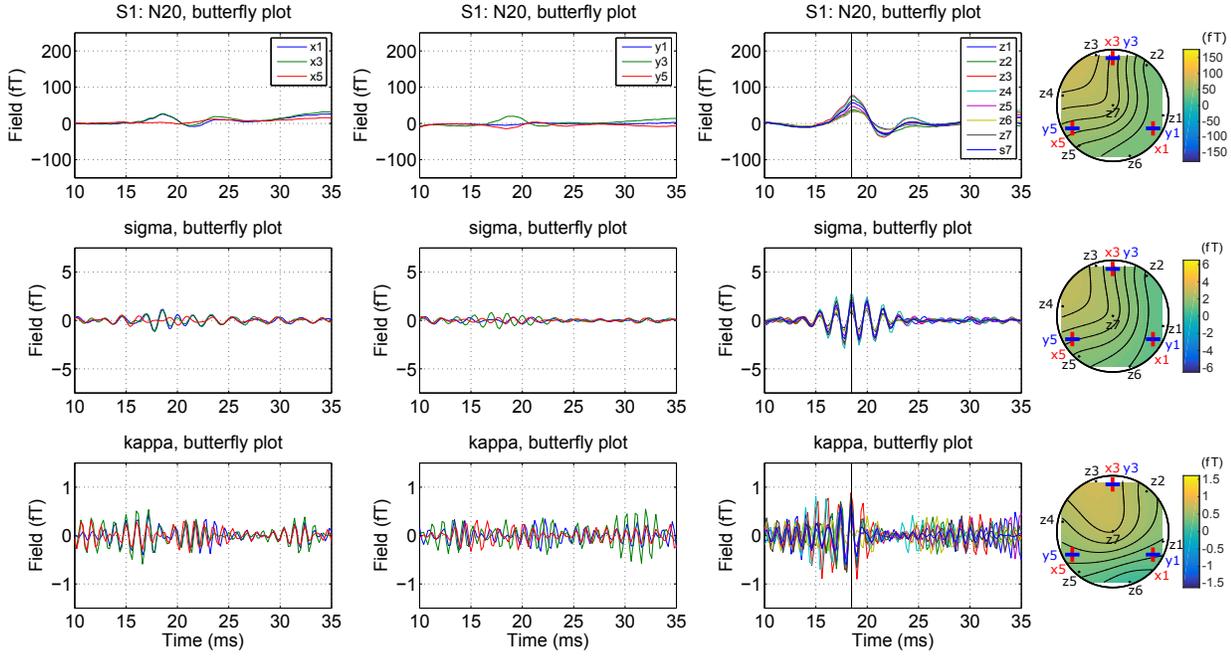}
\caption{\label{Fig:MEG_s1}SEF of N20, $\sigma$-burst and $\kappa$-burst of subject 1 (S1). The solid lines in the $z$-channels plots indicate the time at which the field map is generated. The positions of the $x$- and $y$-sensors are also indicated in the maps.}
\end{figure*}

\begin{figure*}[t]
\includegraphics[width=0.95\textwidth]{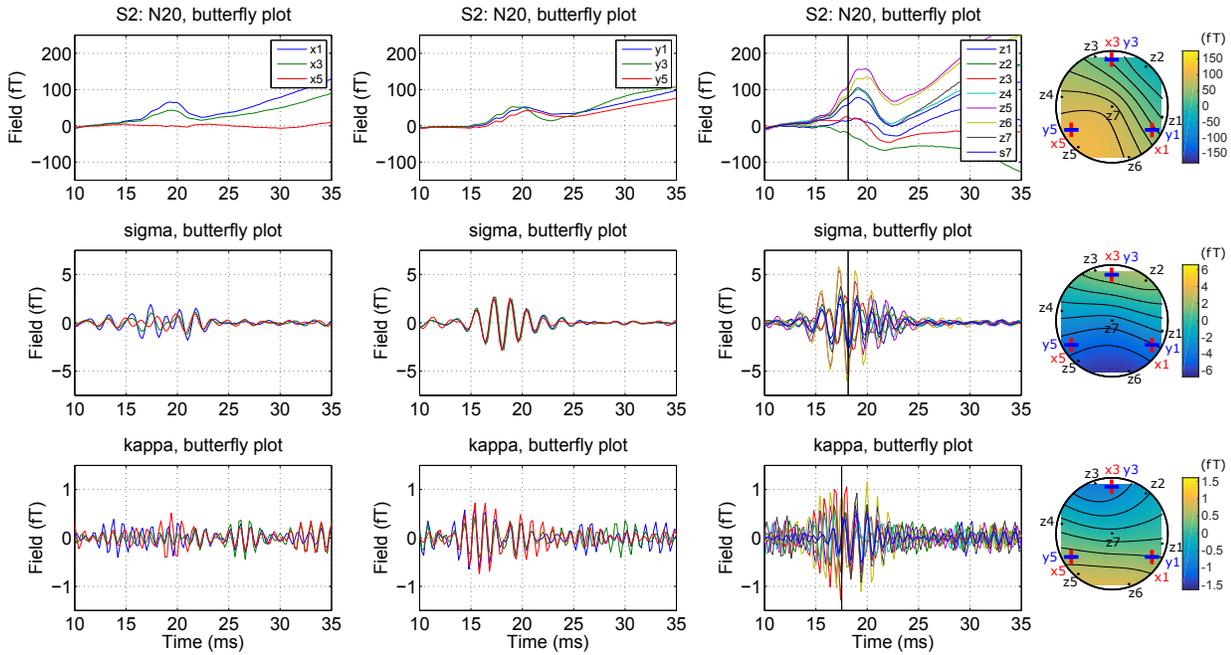}
\caption{\label{Fig:MEG_s2}SEF of N20, $\sigma$-burst and $\kappa$-burst of subject 2 (S2) for all sensors together with the corresponding field maps at 18.2\,ms for the N20 and the $\sigma$-burst. The $\kappa$-burst map has been obtained at 17.6\,ms. The solid lines in the $z$-channels plots indicate the time at which the field map is generated. The positions of the $x$- and $y$-sensors are also indicated in the maps.}
\end{figure*}

 Bandpass digital time-domain filtering of the averaged data in the frequency window from 450\,Hz to 750\,Hz and from 850\,Hz to 1200\,Hz was performed to reveal the so-called $\sigma$- and $\kappa$-bursts, respectively. In \fref{Fig:MEG_s1} and \ref{Fig:MEG_s2}, the N20 and the $\sigma$-burst are shown for subjects~1 (S1) and 2 (S2) using magnetometer data in order to prevent distortions of the $x$-, $y$- and $z$-sensors caused by the non-negligible signals in the reference sensors $x9$, $y9$ and $s9$, respectively. For the $\kappa$-burst, $x9$, $y9$ and $s9$ were used to implement gradiometers due to its negligible signals at the frequency range in question. The magnetic field maps were obtained using the 7 small $z$-sensors in the bottom plane. The data for subject~3 are qualitatively similar to those of S2 and therefore not shown here.

\par Due to the limited number of channels, we did not attempt source localisation. However, from the similarity of the magnetic field maps for S1 we can, for the present discussion, assume that the sources for N20 and $\sigma$-burst co-localise. It is known that the N20 originates from area 3b of the somatosensory cortex SI~\cite{Tanosaki2002}. This can also be assumed for S2, albeit with some relative rotation between the two underlying generators.

\begin{table}[t]
\caption{\label{Tab:SNR} The SNRs for the $\sigma$-burst and the $\kappa$-burst obtained with the bottom $z$-sensors calculated for subjects S1 and S2. The parameter $t$-average($z$1-$z$7) is the SNR of time averaged small sensors $z1$-$z7$.}
\begin{indented}
\item[] \begin{tabular}{@{}lllll}
\br
	sensor& $\sigma$ (S1)& $\sigma$ (S2) &  $\kappa$ (S1)& $\kappa$ (S2)\\
\mr
			$z$1 & 2.88 & 9.06 & 0.99 & 2.25 \\
			$z$2 & 4.25 & 6.34 & 1.45 & 1.85 \\
			$z$3 & 8.42 & 7.35 & 2.71 & 3.49 \\
			$z$4 & 6.61 & 6.94 & 2.08 & 2.00 \\
			$z$5 & 5.01 & 13.2 & 0.85 & 1.92 \\
			$z$6 & 3.35 & 11.9 & 0.78 & 3.66 \\
			$z$7 & 7.35 & 11.9 & 1.57 &  1.99\\
			$t$-average($z$1-$z$7) & 10.2 & 10.6 & 2.31 & 2.93\\
			$s$7 & 13.3 & 10.3 & 2.44 & 2.88 \\
\br
\end{tabular}
\end{indented}
\end{table}

\par The flux density noise $S_{B}^{1/2}$ of the resting brain drops below 0.5\,$\textrm{fT}/\textrm{Hz}^{1/2}$ at frequencies about 400\,Hz when measured over the somatosensory cortex~\cite{Koerber2016}. Consequently, the detection of high-frequency oscillations above this frequency by the large pick-up coil should yield a higher SNR in comparison to the small coils, given a suitable sensor position. In the following we focus the analysis to the bottom $z$-channels as those show the clearest high frequency signals. The SNR for the $\sigma$- and $\kappa$-bursts were determined by taking the ratio of their rms values in the window from 15\,ms to 25\,ms and a noise rms determined for the time window from 25\,ms to 50\,ms. We also calculate the SNR for the time-averaged sensors $z1-z7$. The results are shown in table~\ref{Tab:SNR}. For S1, the position of the module is not central over the N20 generator. Therefore, the zero-field line is not covered so that in this case we obtain a maximum SNR for the large sensor $s7$. In contrast, for S2 the zero-field line is covered, leading to a maximum SNR obtained by $z5$ and not $s7$.

\par For the $\kappa$-burst, this trend is also observed, even though here the maximum SNR for S1 for sensor s7 is not attained. The general low SNR is the likely cause for this and an SNR above 2 is only observed for $z3$, $z4$ and $s7$, respectively. For S2, all the sensors have an SNR close or above 2 and the map at 17.6\,ms appears dipolar-like similar to the $\sigma$-burst. 

\par We briefly discuss the tangential components. If the centre of the module is positioned just above the neuronal source the tangential components should be maximum. However, in our case the additional centre-distance of 11.5\,mm of the $x$- and $y$- sensors to the bottom plane will result in a signal reduction. For S1, the module was not placed over the neuronal generator leading to the $x$ and $y$-components being significantly smaller than the $z$-component. Considering the $\sigma$-burst of S2, the zero-field line lies between $z$7 and the line connecting $z2$ and $z3$, consequently the $y$-sensors sense very similar signals. This is qualitatively also the case for the $\kappa$-burst. Lastly, the $x$-channels show significantly weaker high-frequency component signals.
 
\par We want to point out that the emphasis of this work is not on a detailed spatio-temporal analysis of the various burst signals; this should be addressed in a future study. Instead, we focus on the possibility to record burst activity at around 1\,kHz reliably with sufficient SNR by means of ultra-sensitive multichannel instrumentation as demonstrated here for the first time unambiguously for S2.

\section{\label{sec:conclusion}CONCLUSION}
We presented a detailed description of the design considerations and the performance of our prototype 18-channel module which can be used for e.g. NMR spectroscopy or  MEG. It can be combined with additionally up to 6 modules retaining a regular sensor-grid in all three spatial directions. Deployment in the BMSR-2 allows the use of magnetometer pick-up coils with reference sensors enabling software gradiometers if needed. Since this is to be used as a multi-purpose device, we opted for a hybrid system with overlapping coils of different sizes. This, together with the magnetometer design, represents the best option to achieve maximum SNR, which is a complex function of the origin of noise contributions and source depth. FEM simulations showed that for samples with a few centimetres in dimension, the influence of the Nb tubes, necessary to shield the SQUIDs from pulsed fields, on the homogeneity is comparable with that of a real 1\,m diameter Helmholtz coil system, rendering the prototype module usable for high precision NMR experiments. 
\par We performed successfully MEG and ULF NMR, showing the versatility of the system. We have shown that the potential gain in SNR by deploying different-size sensor loops depends on the dominant source of noise, i.e., environmental or SQUID-intrinsic and the location of the underlying source. With this ultra-sensitive multi-channel system we could detect 1\,kHz MEG components non-invasively in somatosensory evoked activity, a feature thus far only possible by either ultra-sensitive EEG~\cite{Fedelethesis} or with our ultra-low noise single-channel device. The here demonstrated multi-channel detection forms an important step towards a non-invasive localisation of spiking activity by MEG.  

\ack
The authors thank Allard Schnabel, Sassan Ali Valiolahi and Henry Barthelmess for fruitful discussions and contributions. The immaculate machining of the prototype module by Stelter Werkzeugbau, Kummerfeld, Germany is acknowledged. 
\par This  project  has  received  funding  from  the  \textit{European Union's Horizon 2020 research and innovation programme under grant agreement No 686865}.

\section*{References}
\bibliographystyle{iopart-num}
\bibliography{bibliography}

\providecommand{\newblock}{}
\begin{thebibliography}{10}
\expandafter\ifx\csname url\endcsname\relax
  \def\url#1{{\tt #1}}\fi
\expandafter\ifx\csname urlprefix\endcsname\relax\def\urlprefix{URL }\fi
\providecommand{\eprint}[2][]{\url{#2}}

\bibitem{Foglietti1992}
Foglietti V 1992 {Multichannel Instrumentation for Biomagnetism} {\em
  {Superconducting Devices and Their Applications}\/} ed Koch H and L{\"u}bbig
  H (Springer) pp 487--501

\bibitem{Vrba2006}
Vrba J, Nenonen J and Trahms L 2006 Biomagnetism {\em {The SQUID Handbook
  volume II: Applications of SQUIDs and SQUID systems}\/} ed Clarke J and
  Braginski A~I (Wiley-VCH) pp 269--390

\bibitem{Clarke2007}
Clarke J, Hatridge M and M{\"o\ss}le M 2007 {\em Ann. Rev. Biomed. Eng.\/} {\bf
  9} 389--413

\bibitem{ULFNMR2014}
Kraus R~H~J, Espy M~A, Magnelind P~A and Volegov P~L 2014 {\em {Ultra-Low Field
  Nuclear Magnetic Resonance}\/} (Oxford University Press)

\bibitem{Zotev2007}
Zotev V~S, Matlashov A~N, Volegov P~L, Urbaitis A~V, Espy M~A and Jr R~H~K 2007
  {\em Supercond. Sci. Technol.\/} {\bf 20} S367

\bibitem{Vesanen2012}
Vesanen P, Nieminen J, Zevenhoven K, Dabek J, Parkkonen L, Zhdanov A,
  Luomahaara J, Hassel J, Penttil{\"a} J, Simola J, Ahonen A, M{\"a}kel{\"a} J
  and Ilmoniemi R 2013 {\em Magn. Reson. Med.\/} {\bf 69} 1795--1804

\bibitem{Fedele2015}
Fedele T, Scheer H, Burghoff M, Curio G and K{\"o}rber R 2015 {\em Physiol.
  Meas.\/} {\bf 36} 357--368

\bibitem{Koerber2013}
K{\"o}rber R, Nieminen J, H{\"o}fner N, Jazbin{\v{s}}ek V, Scheer H~J, Kim K
  and Burghoff M 2013 {\em J. Magn. Reson.\/} {\bf 237} 182--190

\bibitem{Bork2000}
Bork J, Hahlbohm H~D, Klein R and Schnabel A 2000 {\em Proc. of Biomag\/} {\bf
  2000} 970--973

\bibitem{Ahonen1993a}
Ahonen A~I, H{\"a}m{\"a}l{\"a}inen M~S, Ilmoniemi R~J, Kajola M~J, Knuutila
  J~E~T, Simola J~T and Vilkman V~A 1993 {\em IEEE Trans. Biomed. Eng.\/} {\bf
  40} 859--869

\bibitem{Drung2007}
Drung D, Assmann C, Beyer J, Kirste A, Peters M, Ruede F and Schurig T 2007
  {\em IEEE Trans. Appl. Supercond.\/} {\bf 17} 699--704 ISSN 1051-8223

\bibitem{terBrake1986}
ter Brake H, Fleuren F, Ulfman J and Flokstra J 1986 {\em Cryogenics\/} {\bf
  26} 667--670

\bibitem{Claycomb1999}
Claycomb J~R and Miller J~H 1999 {\em Rev. Sci. Instrum.\/} {\bf 70} 4562--4568

\bibitem{Hilschenz2013}
Hilschenz I, K{\"o}rber R, Scheer H~J, Fedele T, Albrecht H~H, Cassar{\'a} A~M,
  Hartwig S, Trahms L, Haase J and Burghoff M 2013 {\em Magn. Reson. Imaging\/}
  {\bf 31} 171--177

\bibitem{Matlashov2015}
Matlashov A, Magnelind P, Volegov P and Espy M 2015 {\em IEEE Trans. Appl.
  Supercond.\/} {\bf SQ-P09} 1--3

\bibitem{Luomahaara2011}
Luomahaara J, Vesanen P, Penttil{\"a} J, Nieminen J, Dabek J, Simola J,
  Kiviranta M, Gr{\"o}nberg L, Zevenhoven C, Ilmoniemi R and Hassel J 2011 {\em
  Supercond. Sci. Technol.\/} {\bf 24} 075020

\bibitem{Hwang2014}
Hwang S~M, Kim K, Yu K~K, Lee S~J, Shim J~H, K{\"o}rber R and Burghoff M 2014
  {\em Appl. Phys. Lett.\/} {\bf 104} 062602

\bibitem{Magnicon}
See http://www.magnicon.com/squid-electronics/ for description of electronics
  and data sheets.

\bibitem{Clarke2004}
Clarke J and Braginski A~I 2004 {\em The SQUID handbook volume I: Fundamentals
  and technology of SQUIDs and SQUID systems\/} vol~1 (Wiley Online Library)

\bibitem{Nenonen1996}
Nenonen J, Montonen J and Katila T 1996 {\em Rev. Sci. Instrum.\/} {\bf 67}
  2397--2405

\bibitem{Matlashov2012}
Matlashov A, Burmistrov E, Magnelind P, Schultz L, Urbaitis A, Volegov P, Yoder
  J and Espy M 2012 {\em Physica C\/} {\bf 482} 19 -- 26 ISSN 0921-4534

\bibitem{Hartwig2011}
Hartwig S, Voigt J, Scheer H~J, Albrecht H~H, Burghoff M and Trahms L 2011 {\em
  J. Chem. Phys.\/} {\bf 135} 054201

\bibitem{Niedermeyer2004}
Niedermeyer E and Lopes~da Silva F 2004 {\em {Electroencephalography: basic
  principles, clinical applications and related fields.}\/} (Philadelphia:
  Lippincott Williams \& Wilkins)

\bibitem{Tanosaki2002}
Tanosaki M, Suzuki A, Kimura T, Takino R, Haruta Y, Hoshi Y and Hashimoto I
  2002 {\em Neuroreport\/} {\bf 13} 1519--1522

\bibitem{Koerber2016}
K{\"o}rber R, Storm J~H and Seton H 2016 {\em Supercond. Sci. Technol.\/} {\bf
  submitted SUST-101540}

\bibitem{Fedelethesis}
Fedele T 2014 {\em High-frequency electroencephalography (hf-EEG): Non-invasive
  detection of spike-related brain activity\/} Ph.D. thesis Technische
  Universit{\"a}t Berlin

\end{thebibliography}

\end{document}